# Orbital Mechanics near a Rotating Asteroid

Yu Jiang[1, 2], Hexi Baoyin[2]
1. State Key Laboratory of Astronautic Dynamics, Xi'an Satellite Control Center, Xi'an 710043, China
2. School of Aerospace, Tsinghua University, Beijing 100084, China
Y. Jiang (✉) e-mail: jiangyu_xian_china@163.com (corresponding author)

**Abstract.** This study investigates the different novel forms of the dynamical equations of a particle orbiting a rotating asteroid and the effective potential, the Jacobi integral, etc. on different manifolds. Nine new forms of the dynamical equations of a particle orbiting a rotating asteroid are presented, and the classical form of the dynamical equations has also been found. The dynamical equations with the potential and the effective potential in scalar form in the arbitrary body-fixed frame and the special body-fixed frame are presented and discussed. Moreover, the simplified forms of the effective potential and the Jacobi integral have been derived. The dynamical equation in coefficient-matrix form has been derived. Other forms of the dynamical equations near the asteroid are presented and discussed, including the Lagrange form, the Hamilton form, the symplectic form, the Poisson form, the Poisson-bracket form, the cohomology form, and the dynamical equations on the Kähler manifold and another complex manifold. Novel forms of the effective potential and the Jacobi integral are also presented. The dynamical equations in scalar form and coefficient-matrix form can aid in the study of the dynamical system, the bifurcation, and the chaotic motion of the orbital dynamics of a particle near a rotating asteroid. The dynamical equations of a particle near a rotating asteroid are presented on several manifolds, including the symplectic manifold, the Poisson manifold, and complex manifolds, which may lead to novel methods of studying the motion of a particle in the potential field of a rotating asteroid.

## 1 Introduction

The orbital mechanics of a spacecraft around an asteroid has recently become a topic of interest. If the spacecraft is sufficiently far from the asteroid, the asteroid can be approximately modelled as a particle, and the solar gravitation must be considered to study the dynamics of the spacecraft near the asteroid. If the shape of the body is nearly a spheroid, as in the case of Earth, Mars, etc., the classical method of the Legendre polynomial series can be applied to study the gravitational potential of the bodies, and the convergence of the series is sufficiently fast (Riaguas et al. 1999), and the stability regions for orbital motion in uniformly rotating second degree and order gravity fields can be calculated numerically (Hu and Scheeres 2004). When the shape of the asteroid is irregular, the Legendre polynomial series barely converges in the vicinity of the asteroid (Riaguas et al. 1999; Blesa. 2006); it may perhaps diverge at certain points in the gravitational field of the asteroid (Balmino. 1994; Elipe et al. 2003).

The motion of a particle near some other simply shaped bodies has been simulated for the purpose of studying the chaotic motion, periodic orbits, equilibrium points, etc. around such asteroids (Scheeres 1994; Riaguas et al. 1999; Hu and Scheeres 2002; Elipe et al. 2003; Broucke and Elipe. 2005; Garcia-Abdeslem 2005; Mufti 2006a, 2006b; Alberti and Vidal. 2007; Fukushima. 2010; Liu et al. 2011; Chappell et al. 2013). For instance, some asteroids can be found for which an elongated body is the main feature of their shape, e.g., Eros, Ida, and Amaltea; this elongated shape means that the pseudo-spherical approach to the gravitational field of such a mass distribution is far from providing a sufficient description of the true effect of the gravitational field of the asteroid (Riaguas et al. 1999). Other specially shaped bodies have also been considered for the analysis of the motion in their gravitational fields. Blesa (2006) calculated the gravitational potentials of square and triangular plates and presented several families of periodic orbits in the planes of these plates by computing the Poincaré surface. Broucke and Elipe (2005) studied the dynamics and the families of periodic orbits in a potential field of a solid circular ring. Alberti and Vidal (2007) discussed the motion of a particle moving in the gravitational field induced by a homogeneous annular disk on a fixed plane. Fukushima (2010) described a precise numerical method to evaluate the acceleration vector of the gravitational force caused by a uniform ring or disk. Liu et al. (2011) investigated the equilibria of a rotating homogeneous cube and the periodic orbits around these equilibria. These researches provide some simple and important examples of orbital dynamics in the potential field of a rotating body, and they can help one to understand the dynamics of orbits near an irregularly shaped rotating asteroid.

Werner (1994) modelled the geometries of irregularly shaped asteroids with a constant-density polyhedron, expressed the exterior gravitational potential and acceleration components in terms of the polyhedron's edges and vertex angles, and applied this method to the inner Martian satellite Phobos with 146 vertices and 288 triangular faces. Werner and Scheeres (1996) derived the exterior gravitation of a constant-density polyhedron in closed form. Scheeres et al. (1996) used a radar-derived physical model of the asteroid 4769 Castalia to study the orbital dynamics near the asteroid with a kilometre-sized particle and established a Jacobi integral for such a particle in the vicinity of the asteroid; the zero-velocity surfaces were discussed, and the periodic orbit families were computed. After this, several studies were conducted concerning the dynamics near other asteroids, such as 4179 Toutatis (Scheeres et al. 1998), 433 Eros (Scheeres et al. 2000), 216 Kleopatra (Yu &

Baoyin. 2012a, 2012b), as well as asteroids 2002 AT4 and 1989 ML (Scheeres 2012).

As an extension to the study about satellite orbits around an asteroid under the assumption of Newtonian gravity, the general relativistic corrections can be calculated (Iorio 2005). However, these effects will only become significant for special situations, such as those involving high orbital speeds or a fast spinning rate for the central body producing the Lense-Thirring effect (Iorio 2005; Iorio et al. 2011).

The polyhedron method can effectively avoid poor convergence behaviour when modelling an irregular gravitational field. This method makes it possible to combine theoretical research and higher-precision numerical methods to discuss the dynamics of orbits close to certain special asteroids, perhaps including the periodic orbit, the quasi-periodic orbits, the equilibrium points, the dynamical system, the bifurcation, and the chaotic motion of the orbital dynamics of a particle near a rotating asteroid. Theoretical research of the orbital mechanics near a rotating asteroid is therefore required in certain special forms; these special forms include the dynamical equations in the scalar form, the coefficient-matrix form, and the Lagrange form. The dynamical equations of a particle near a rotating asteroid on several different novel manifolds may perhaps lead to some novel methods of studying the motion of a particle in the potential field of a rotating body; these manifolds include the symplectic manifold, the Poisson manifold, and complex manifolds.

In this work, we are interested in the different novel forms of the dynamical equations of a particle around a rotating asteroid and in the effective potential, the Jacobi integral, etc. on different manifolds. In addition to the classical form of the dynamical equations, 9 new forms of the dynamical equations of a particle near a rotating asteroid are presented: the scalar form, the coefficient-matrix form, the Lagrange form, the Hamilton form, the symplectic form, the Poisson-bracket form, the Poisson form, the complex form and the cohomology form. We proceed as follows: We first discuss the classical form of the equations of motion of a particle around a rotating asteroid. Then, the dynamical equations, along with the potential and the effective potential, in scalar form in an arbitrary body-fixed frame and a special body-fixed frame are presented and studied. The simplified forms of the effective potential and the Jacobi integral are given. The dynamical equations in coefficient-matrix form are derived, and this form of the dynamical equations is in the form of a first-order ordinary differential equation; using dynamical systems theory, the dynamical equations in scalar form and coefficient-matrix form can aid in the study of the dynamical system, the bifurcation, and the chaotic motion of the orbital dynamics of a particle around a rotating asteroid.

Other forms of the dynamical equations near the asteroid are presented and studied, including the Lagrange form, the Hamilton form, the symplectic form, the Poisson-bracket form, the Poisson form, the cohomology form, and the dynamical equations on the Kähler manifold and another complex manifold. Novel forms of the effective potential and the Jacobi integral are also presented.

The dynamical equation in the classical form is the motion of the particle expressed in 3-dimensional Euclidean space, and the dynamical equation in Hamilton form is the motion of the particle expressed in phase space. Analogously, the dynamical equations in symplectic form and Poisson-bracket form represent the motion of the particle expressed on the symplectic manifold, while the dynamical equation in Poisson form is the motion of the particle expressed on the Poisson manifold. The motion of a particle in the potential field of a rotating asteroid can also be expressed on complex manifolds, including the Kähler manifold. Using the Hodge star operator and the differential operator, the dynamical equation of the particle can be expressed simply and beautifully in the cohomology form.

## 2 Equations of Motion near a Rotating Asteroid

As stated at the end of Section 1, in addition to the classical form of the dynamical equations, we present 9 new forms of the dynamical equations of a particle near a rotating asteroid. We begin with the classical form.

### 2.1 Classical Form

The equation of motion can be expressed as a second-order ordinary differential equation (Scheeres et al. 1996)

$$\ddot{\mathbf{r}}+2\boldsymbol{\omega}\times\dot{\mathbf{r}}+\boldsymbol{\omega}\times(\boldsymbol{\omega}\times\mathbf{r})+\dot{\boldsymbol{\omega}}\times\mathbf{r}+\frac{\partial U(\mathbf{r})}{\partial\mathbf{r}}=0, \tag{1}$$

where $\mathbf{r}$ is the radius vector from the asteroid's centre of mass to the particle, the first and second time derivatives of $\mathbf{r}$ are with respect to the body-fixed coordinate system, $\boldsymbol{\omega}$ is the rotational angular velocity vector of the asteroid relative to inertial space, and $U(\mathbf{r})$ is the gravitational potential of the asteroid. Eq. (1) is the equation of motion of the particle expressed in 3-dimensional Euclidean space. The Jacobi integral can be given as (Scheeres et al. 1996)

$$H=\frac{1}{2}\dot{\mathbf{r}}\cdot\dot{\mathbf{r}}-\frac{1}{2}(\boldsymbol{\omega}\times\mathbf{r})\cdot(\boldsymbol{\omega}\times\mathbf{r})+U(\mathbf{r}), \tag{2}$$

which is the Hamilton function of the dynamical system and is the integral of the relative energy. If $\boldsymbol{\omega}$ is time invariant, then $H$ is also time invariant and is referred to as the Jacobi constant. We can write

$$E=\frac{1}{2}\mathbf{v}_I\cdot\mathbf{v}_I+U(\mathbf{r}), \tag{3}$$

where $\mathbf{v}_I=\dot{\mathbf{r}}+\boldsymbol{\omega}\times\mathbf{r}$ is the velocity of the particle relative to inertial space, and $E$ is the mechanical energy of the particle. Then, if $\boldsymbol{\omega}$ is time invariant, the mechanical energy $E$ is not conservative, but the Jacobi integral is conservative.

The kinetic energy is then

$$T = \frac{1}{2}\left(\dot{\mathbf{r}} + \boldsymbol{\omega} \times \mathbf{r}\right) \cdot \left(\dot{\mathbf{r}} + \boldsymbol{\omega} \times \mathbf{r}\right). \tag{4}$$

The effective potential is defined as (Yu et al. 2012a)

$$V\left(\mathbf{r}\right) = -\frac{1}{2}\left(\boldsymbol{\omega} \times \mathbf{r}\right) \cdot \left(\boldsymbol{\omega} \times \mathbf{r}\right) + U\left(\mathbf{r}\right). \tag{5}$$

Substituting this into Eq. (1) yields

$$\ddot{\mathbf{r}} + 2\boldsymbol{\omega} \times \dot{\mathbf{r}} + \dot{\boldsymbol{\omega}} \times \mathbf{r} + \frac{\partial V\left(\mathbf{r}\right)}{\partial \mathbf{r}} = 0. \tag{6}$$

The Hamilton function can then be written as (Yu et al. 2012a)

$$H = \frac{1}{2}\dot{\mathbf{r}} \cdot \dot{\mathbf{r}} + V\left(\mathbf{r}\right). \tag{7}$$

The orbital dynamics of a particle near a variable-velocity rotating asteroid are markedly different than the dynamics of a particle near the uniformly rotating asteroid (Scheeres et al. 1996, 1998). In particular, if $\boldsymbol{\omega}$ is time invariant, then Eq. (6) can be expressed as (Yu et al. 2012a)

$$\ddot{\mathbf{r}} + 2\boldsymbol{\omega} \times \dot{\mathbf{r}} + \frac{\partial V\left(\mathbf{r}\right)}{\partial \mathbf{r}} = 0, \tag{8}$$

and Eq. (1) can be expressed as (Yu et al. 2013)

$$\ddot{\mathbf{r}} + 2\boldsymbol{\omega} \times \dot{\mathbf{r}} + \boldsymbol{\omega} \times \left(\boldsymbol{\omega} \times \mathbf{r}\right) + \frac{\partial U\left(\mathbf{r}\right)}{\partial \mathbf{r}} = 0. \tag{9}$$

The zero-velocity manifolds can be defined by the following equation (Scheeres et al. 1996; Yu et al. 2012a):

$$V\left(\mathbf{r}\right) = H. \tag{10}$$

The forbidden region for the particle can be determined from the inequality $V\left(\mathbf{r}\right) > H$, while the allowed region for the particle can be determined from the inequality $V(\mathbf{r}) < H$. The equation $V(\mathbf{r}) = H$ implies that the particle is static relative to the rotating body-fixed frame.

## 2.2 Dynamical Equations in Scalar Form

### 2.2.1 General Dynamical Equations in the Arbitrary Body-fixed Frame

The body-fixed frame is defined by an orthonormal right-handed set of unit vectors $\{\mathbf{e}\}$:

$$\{\mathbf{e}\} \equiv \begin{Bmatrix} \mathbf{e}_{\mathrm{x}} \\ \mathbf{e}_{\mathrm{y}} \\ \mathbf{e}_{\mathrm{z}} \end{Bmatrix}. \tag{11}$$

The dynamical equations of the particle can be written as

$$\begin{cases} \ddot{x} + \dot{\omega}_y z - \dot{\omega}_z y + 2\omega_y \dot{z} - 2\omega_z \dot{y} + \omega_x \omega_y y - \omega_y^2 x - \omega_z^2 x + \omega_z \omega_x z + \dfrac{\partial U}{\partial x} = 0 \\ \ddot{y} + \dot{\omega}_z x - \dot{\omega}_x z + 2\omega_z \dot{x} - 2\omega_x \dot{z} + \omega_y \omega_z z - \omega_z^2 y - \omega_x^2 y + \omega_x \omega_y x + \dfrac{\partial U}{\partial y} = 0 \\ \ddot{z} + \dot{\omega}_x y - \dot{\omega}_y x + 2\omega_x \dot{y} - 2\omega_y \dot{x} + \omega_x \omega_z x - \omega_x^2 z - \omega_y^2 z + \omega_y \omega_z y + \dfrac{\partial U}{\partial z} = 0 \end{cases}. \tag{12}$$

Using the effective potential, the dynamical equations can be rewritten as

$$\begin{cases} \ddot{x} + \dot{\omega}_y z - \dot{\omega}_z y + 2\omega_y \dot{z} - 2\omega_z \dot{y} + \dfrac{\partial V}{\partial x} = 0 \\ \ddot{y} + \dot{\omega}_z x - \dot{\omega}_x z + 2\omega_z \dot{x} - 2\omega_x \dot{z} + \dfrac{\partial V}{\partial y} = 0 \\ \ddot{z} + \dot{\omega}_x y - \dot{\omega}_y x + 2\omega_x \dot{y} - 2\omega_y \dot{x} + \dfrac{\partial V}{\partial z} = 0 \end{cases}. \tag{13}$$

Eq. (12) and Eq. (13) are the general dynamical equations of a particle near a rotating asteroid expressed in the arbitrary body-fixed frame.

### 2.2.2 General Dynamical Equations in the Special Body-fixed Frame

Let $\omega$ be the norm of the vector $\boldsymbol{\omega}$. If the unit vector $\mathbf{e}_{\mathbf{z}}$ is defined by $\boldsymbol{\omega} = \omega \mathbf{e}_{\mathbf{z}}$, then the dynamical equations in component form simplify to

$$\begin{cases} \ddot{x} - \dot{\omega} y - 2\omega \dot{y} - \omega^2 x + \dfrac{\partial U}{\partial x} = 0 \\ \ddot{y} + \dot{\omega} x + 2\omega \dot{x} - \omega^2 y + \dfrac{\partial U}{\partial y} = 0 \\ \ddot{z} + \dfrac{\partial U}{\partial z} = 0 \end{cases}. \tag{14}$$

The effective potential becomes

$$V = U - \frac{\omega^2}{2}\left(x^2 + y^2\right); \tag{15}$$

it is related only to the position of the particle in the body-fixed frame and is independent of $z$.

The Jacobi integral reduces to

$$H = U + \frac{1}{2}\left(\dot{x}^2 + \dot{y}^2 + \dot{z}^2\right) - \frac{\omega^2}{2}\left(x^2 + y^2\right) \quad . \tag{16}$$

If $\omega$ is time invariant, then $H$ is a constant, meaning that the integral of the relative energy is conserved.

The Lagrange function is given by

$$L = \frac{1}{2}\left(\dot{x}^2 + \dot{y}^2 + \dot{z}^2\right) + \frac{1}{2}\omega^2\left(x^2 + y^2\right) + \omega\left(x\dot{y} - \dot{x}y\right) - U . \tag{17}$$

Using the effective potential, the dynamical equations can be written as

$$\begin{cases} \ddot{x} - \dot{\omega} y - 2\omega \dot{y} + \dfrac{\partial V}{\partial x} = 0 \\ \ddot{y} + \dot{\omega} x + 2\omega \dot{x} + \dfrac{\partial V}{\partial y} = 0 \\ \ddot{z} + \dfrac{\partial V}{\partial z} = 0 \end{cases}. \tag{18}$$

The effective potential has the following properties:

a) If $|\mathbf{r}| \to +\infty$, $V(\mathbf{r}) \to -\dfrac{\omega^2}{2}\left(x^2 + y^2\right)$.

b) $$\begin{cases} \dfrac{\partial V(\mathbf{r})}{\partial x} = -\omega^2 x + \dfrac{\partial U(\mathbf{r})}{\partial x} \\ \dfrac{\partial V(\mathbf{r})}{\partial y} = -\omega^2 y + \dfrac{\partial U(\mathbf{r})}{\partial y} \\ \dfrac{\partial V(\mathbf{r})}{\partial z} = \dfrac{\partial U(\mathbf{r})}{\partial z} \end{cases}.$$

c) The asymptotic surface of $V = V(\mathbf{r})$ is a circular cylindrical surface that can be expressed as $V^* = -\frac{\omega^2}{2}(x^2 + y^2)$; the radius of the circular cylindrical surface is $\frac{\sqrt{2}}{2}\omega\sqrt{x^2+y^2}$ .

d) The function $V = V(\mathbf{r})$ is a 3-dimensional smooth manifold, and $V(\mathbf{r}) = C$ denotes a 2-dimensional curved surface, where $C$ is a constant.

### 2.2.3 Dynamical Equations Near the Uniformly Rotating Asteroid in the Special Body-fixed Frame

If the unit vector $\mathbf{e_z}$ is defined by $\boldsymbol{\omega} = \omega \mathbf{e_z}$, for the uniformly rotating asteroid, the dynamical equations in scalar form can be expressed as (Scheeres et al. 1996)

$$\begin{cases} \ddot{x} - 2\omega\dot{y} - \omega^2 x + \dfrac{\partial U}{\partial x} = 0 \\ \ddot{y} + 2\omega\dot{x} - \omega^2 y + \dfrac{\partial U}{\partial y} = 0 \\ \ddot{z} + \dfrac{\partial U}{\partial z} = 0 \end{cases}, \tag{19}$$

when the effective potential, the Jacobi integral, and the Lagrange function are in the forms of Eq. (15) - Eq. (17).

Using the effective potential, the dynamical equations of a particle near the uniformly rotating asteroid can be rewritten as

$$\begin{cases} \ddot{x} - 2\omega\dot{y} + \frac{\partial V}{\partial x} = 0 \\ \ddot{y} + 2\omega\dot{x} + \frac{\partial V}{\partial y} = 0 \\ \ddot{z} + \frac{\partial V}{\partial z} = 0 \end{cases}. \tag{20}$$

## 2.3 Dynamical Equations in Coefficient-Matrix Form

If the dynamical equations of the particle can be expressed in the coefficient-matrix form, dynamical systems theory can be easily applied to study the orbital dynamical system, the bifurcation, and the chaotic motion of the orbital motion of a particle near a rotating asteroid.

Let $\mathbf{v} = \dot{\mathbf{r}}$ and $\boldsymbol{\tau} = \dot{\boldsymbol{\omega}}$, where $\hat{\boldsymbol{\omega}} = \begin{pmatrix} 0 & -\omega_z & \omega_y \\ \omega_z & 0 & -\omega_x \\ -\omega_y & \omega_x & 0 \end{pmatrix}$ and $\hat{\boldsymbol{\tau}} = \begin{pmatrix} 0 & -\tau_z & \tau_y \\ \tau_z & 0 & -\tau_x \\ -\tau_y & \tau_x & 0 \end{pmatrix}$.

Substituting this into Eq. (6) yields

$$\begin{cases} \dot{\mathbf{r}} = \mathbf{v} \\ \dot{\mathbf{v}} = -\nabla V(\mathbf{r}) - 2\hat{\boldsymbol{\omega}}\mathbf{v} - \hat{\boldsymbol{\tau}}\mathbf{r} \end{cases}. \tag{21}$$

Let $\mathbf{X} = \begin{bmatrix} \mathbf{r} \\ \mathbf{v} \end{bmatrix}$, $\mathbf{A} = \begin{pmatrix} \mathbf{I}_{3\times3} & \mathbf{0}_{3\times3} \\ -2\hat{\boldsymbol{\omega}} & -\hat{\boldsymbol{\tau}} \end{pmatrix}$ and $\mathbf{B} = \begin{pmatrix} \mathbf{0}_{3\times3} \\ -\nabla V(\mathbf{r}) \end{pmatrix}$. Then, Eq. (21) can be written as

$$\dot{\mathbf{X}} = \mathbf{A}\mathbf{X} + \mathbf{B}(\mathbf{X}), \tag{22}$$

where $\mathbf{B} = \mathbf{B}(\mathbf{X})$ is a function of $\mathbf{X}$ and $\boldsymbol{\omega}$; to be precise, $\mathbf{B}$ is a function of $\mathbf{r}$ and $\boldsymbol{\omega}$ and is independent of $\mathbf{v}$. Eq. (22) is the dynamical equations expressed in coefficient-matrix form, and it appears in the form of a first-order ordinary differential equation.

If we define $\mathbf{F}(\mathbf{X})=\mathbf{AX}+\mathbf{B}(\mathbf{X})$, then

$$\frac{d\mathbf{F}(\mathbf{X})}{d\mathbf{X}}=\mathbf{A}+\begin{pmatrix}\mathbf{0}_{3\times3} & \mathbf{0}_{3\times3}\\ -\nabla^2V(\mathbf{r}) & \mathbf{0}_{3\times3}\end{pmatrix}=\begin{pmatrix}\mathbf{I}_{3\times3} & \mathbf{0}_{3\times3}\\ -2\hat{\boldsymbol{\omega}}-\nabla^2V(\mathbf{r}) & -\hat{\boldsymbol{\tau}}\end{pmatrix}. \tag{23}$$

## 2.4 Dynamical Equations in Lagrange Form

The generalised momentum is $\mathbf{p}=(\dot{\mathbf{r}}+\boldsymbol{\omega}\times\mathbf{r})$, and the generalised coordinate is $\mathbf{q}=\mathbf{r}$. The Lagrange function is then given by

$$L=\frac{1}{2}(\dot{\mathbf{r}}+\boldsymbol{\omega}\times\mathbf{r})\cdot(\dot{\mathbf{r}}+\boldsymbol{\omega}\times\mathbf{r})-U(\mathbf{r})=\frac{\mathbf{p}\cdot\mathbf{p}}{2}-U(\mathbf{q}), \tag{24}$$

$$L=\frac{1}{2}\dot{\mathbf{r}}\cdot\dot{\mathbf{r}}+\dot{\mathbf{r}}\cdot(\boldsymbol{\omega}\times\mathbf{r})-V(\mathbf{r})=\frac{1}{2}\dot{\mathbf{q}}\cdot\dot{\mathbf{q}}+\dot{\mathbf{q}}\cdot(\boldsymbol{\omega}\times\mathbf{q})-V(\mathbf{q}), \tag{25}$$

and the dynamical equations of a particle in the potential field of a rotating asteroid can be expressed in the Lagrange form (Libermann & Marle. 1987; Berndt. 1998)

$$\frac{d}{dt}\left(\frac{\partial L}{\partial\dot{\mathbf{q}}}\right)=\frac{\partial L}{\partial\mathbf{q}}. \tag{26}$$

The effective potential is then given by

$$V(\mathbf{q})=-\frac{1}{2}(\boldsymbol{\omega}\times\mathbf{q})\cdot(\boldsymbol{\omega}\times\mathbf{q})+U(\mathbf{q}), \tag{27}$$

and the Hamilton function can be expressed as

$$H=-\frac{\mathbf{p}\cdot\mathbf{p}}{2}+U(\mathbf{q})+\mathbf{p}\cdot\dot{\mathbf{q}}. \tag{28}$$

The zero-velocity manifold can be written as follows:

$$V(\mathbf{q})=-\frac{\mathbf{p}\cdot\mathbf{p}}{2}+U(\mathbf{q})+\mathbf{p}\cdot\dot{\mathbf{q}}. \tag{29}$$

There is no explicit time in Eq. (29).

When the dynamical equations of a particle in the potential field of a rotating asteroid are expressed in the Lagrange form, one can think of the motion as the autonomous curve of the functional $\Phi = \int_{t_0}^{t_1} L dt$.

## 2.5 Symplectic Manifold and Dynamical Equations

The symplectic structure is a differential exterior 2-form $\Omega$ defined on a differentiable manifold $M$; $(M,\Omega)$ is then a symplectic manifold (Fomenko. 1988; Sternberg. 2012). The symplectic manifold $(M,\Omega)$ is an even-dimensional manifold.

### 2.5.1 Dynamical Equations in Hamilton Form

According to Libermann & Marle (1987), a Hamilton system is a triplet $(M,\Omega,H)$, where $(M,\Omega)$ is a symplectic manifold, and $H$ is a real differentiable function defined on $M$; the symplectic manifold $(M,\Omega)$ is referred to as the phase space of the system, and the real differentiable function $H$ is referred to as the Hamilton function of the system.

For the orbital mechanics of a particle near an asteroid, the Hamilton function is

$$\begin{aligned} H &= \frac{m}{2}\dot{\mathbf{r}}\cdot\dot{\mathbf{r}} - \frac{m}{2}(\dot{\mathbf{r}}+\boldsymbol{\omega}\times\mathbf{r})\cdot(\dot{\mathbf{r}}+\boldsymbol{\omega}\times\mathbf{r}) + U(\mathbf{r}) \\ &= -\frac{\mathbf{p}\cdot\mathbf{p}}{2m} + U(\mathbf{q}) + \mathbf{p}\cdot\dot{\mathbf{q}} \end{aligned}. \tag{30}$$

The dynamical equations of a particle in the potential field of a rotating asteroid can be expressed in the Hamilton form (Libermann & Marle. 1987; Berndt. 1998)

$$\begin{cases} \dot{\mathbf{p}} = -\dfrac{\partial H}{\partial \mathbf{q}} \\ \dot{\mathbf{q}} = \dfrac{\partial H}{\partial \mathbf{p}} \end{cases}. \tag{31}$$

If $\boldsymbol{\omega}$ is time invariant, then the asteroid rotates uniformly, and if the Jacobi integral

$H$ is also time invariant, then the symplectic geometric algorithm can be used to calculate the dynamical equations of Eq. (31); otherwise, the asteroid has a variable rotating velocity, and the Jacobi integral $H$ is not a constant. Eq. (31) expresses the dynamical equations of the particle in phase space.

### 2.5.2 Dynamical Equations in Symplectic Form

Define

$$\mathbf{z} = \begin{bmatrix} \mathbf{p} & \mathbf{q} \end{bmatrix}^T, \tag{32}$$

where $\mathbf{z}$ is a $6 \times 1$ vector. Then, the dynamical equations of a particle in the potential field of a rotating asteroid can be expressed in the symplectic form (Marsden & Ratiu. 1999)

$$\dot{\mathbf{z}} = \begin{pmatrix} \mathbf{0} & -\mathbf{I} \\ \mathbf{I} & \mathbf{0} \end{pmatrix} \nabla H(\mathbf{z}), \tag{33}$$

where $\mathbf{I}$ and $\mathbf{0}$ are $3 \times 3$ matrices, and $\nabla H(\mathbf{z}) = \begin{pmatrix} \dfrac{\partial H}{\partial \mathbf{p}} & \dfrac{\partial H}{\partial \mathbf{q}} \end{pmatrix}^T$ is the gradient of $H(\mathbf{z})$. Define $\mathbf{J} = \begin{pmatrix} \mathbf{0} & -\mathbf{I} \\ \mathbf{I} & \mathbf{0} \end{pmatrix}$, where $\mathbf{J}$ is a symplectic matrix, $\mathbf{J}\nabla H(\mathbf{z})$ is the Hamiltonian vector field on the symplectic manifold, and the dynamical equations in symplectic form can be rewritten as

$$\mathbf{J}\dot{\mathbf{z}} + \nabla H(\mathbf{z}) = 0. \tag{34}$$

If $\boldsymbol{\omega}$ is time invariant, then the integral of the relative energy $H(\mathbf{z})$ is also time invariant, and the symplectic geometric algorithm can be used to calculate the dynamical equations Eq. (33) or Eq. (34).

## 2.6 Poisson Bracket and Dynamical Equations

The Poisson bracket $\{f,g\}$ is a bilinear map of two smooth functions $f$ and $g$ on a symplectic manifold $(M,\Omega)$, from $C^{\infty}(M,R)\times C^{\infty}(M,R)$ to $C^{\infty}(M,R)$, which satisfies the following conditions: a) $\{f,g\}$ is skew-symmetric, $\{f,g\}=-\{g,f\}$; and b) the Jacobi identity $\{f,\{g,h\}\}+\{g,\{h,f\}\}+\{h,\{f,g\}\}=0$ is satisfied for any $f,g$ and $h$ (Fomenko. 1988). Using the Poisson bracket, the dynamical equations of a particle in the potential field of a rotating asteroid can be given in Poisson-bracket form on the symplectic manifold $(M,\Omega)$.

Define the Poisson bracket as

$$\{f,g\}=\frac{\partial f}{\partial \mathbf{q}}\cdot\frac{\partial g}{\partial \mathbf{p}}-\frac{\partial f}{\partial \mathbf{p}}\cdot\frac{\partial g}{\partial \mathbf{q}}. \tag{35}$$

The dynamical equations of a particle in the potential field of a rotating asteroid can then be expressed as (Libermann & Marle. 1987)

$$\dot{\mathbf{f}}=\{\mathbf{f},H\}, \tag{36}$$

where the integral of the relative energy is

$$H=-\frac{\mathbf{p}\cdot\mathbf{p}}{2m}+U(\mathbf{q})+\mathbf{p}\cdot\dot{\mathbf{q}}=\frac{1}{2}(\mathbf{p}-\boldsymbol{\omega}\times\mathbf{q})\cdot(\mathbf{p}-\boldsymbol{\omega}\times\mathbf{q})+V(\mathbf{q}). \tag{37}$$

Substituting $\mathbf{p}=\mathbf{f}$ into Eq. (36), one can obtain $\dot{\mathbf{p}}=-\frac{\partial H}{\partial \mathbf{q}}$, while substituting $\mathbf{q}=\mathbf{f}$ into Eq. (36), one can obtain $\dot{\mathbf{q}}=\frac{\partial H}{\partial \mathbf{p}}$.

## 2.7 Poisson Manifold and Dynamical Equations

A Poisson manifold is a differentiable manifold that defines a Poisson structure, which is a bilinear map from $C^{\infty}(M,R)\times C^{\infty}(M,R)$ to $C^{\infty}(M,R)$, and the bilinear map satisfies the following conditions: a) $\{f,g\}$ is skew-symmetric, $\{f,g\}=-\{g,f\}$; b) the Jacobi identity $\{f,\{g,h\}\}+\{g,\{h,f\}\}+\{h,\{f,g\}\}=0$ is satisfied for any $f,g$ and $h$; and c) it is a derivation on each of its arguments, $\{fg,h\}=f\{g,h\}+g\{f,h\}$ (Fomenko. 1988; Sternberg. 2012). Every symplectic manifold is a Poisson manifold, but not every Poisson manifold is a symplectic manifold. $Z$ denotes the manifold formed by $\mathbf{z}=[\mathbf{p}\quad \mathbf{q}]^T$; $Z^*$ is the dual space of $Z$.

We can write

$$\begin{aligned} H &= \frac{1}{2}(\mathbf{p}-\boldsymbol{\omega}\times\mathbf{q})\cdot(\mathbf{p}-\boldsymbol{\omega}\times\mathbf{q})+V(\mathbf{q})=\frac{1}{2}\dot{\mathbf{r}}\cdot\dot{\mathbf{r}}+V(\mathbf{r}) \\ &=\frac{1}{2}\dot{\mathbf{r}}\cdot\dot{\mathbf{r}}-\frac{1}{2}(\boldsymbol{\omega}\times\mathbf{r})\cdot(\boldsymbol{\omega}\times\mathbf{r})+U(\mathbf{r}) \end{aligned}. \tag{38}$$

Define $\Omega^{\#}:Z^*\to Z:\nabla H\to X_H$ as

$$\nabla H(\mathbf{z})=\begin{pmatrix}\dfrac{\partial H}{\partial \mathbf{p}} & \dfrac{\partial H}{\partial \mathbf{q}}\end{pmatrix}^T \to \left(-\frac{\partial H}{\partial \mathbf{q}},\frac{\partial H}{\partial \mathbf{p}}\right)^T, \tag{39}$$

where $\left(-\dfrac{\partial H}{\partial \mathbf{q}},\dfrac{\partial H}{\partial \mathbf{p}}\right)^T \triangleq X_H(\mathbf{z})$, and $X_H$ is the entire smooth Hamiltonian vector field.

$\Gamma(X_H)$ denotes the set of the entire smooth Hamiltonian vector field. The mapping of Eq. (39) can be rewritten as (Marsden & Ratiu. 1999)

$$X_H(\mathbf{z})=\Omega^{\#}\circ\nabla H(\mathbf{z})=\Omega^{\#}\nabla H(\mathbf{z}), \tag{40}$$

where $\circ$ denotes the composition operator of $\Omega^{\#}$ and $\nabla$; it can be omitted without ambiguity. Then, the dynamical equations of a particle in the potential field of a rotating asteroid can be expressed on the Poisson manifold as

$$\dot{\mathbf{z}}=X_H(\mathbf{z}) \tag{41}$$

or

$$\dot{\mathbf{z}} = \Omega^{\#}\nabla H\left(\mathbf{z}\right). \tag{42}$$

Let $\Omega^{\mathrm{b}}$ be the inverse mapping of $\Omega^{\#}$, $\Omega^{\mathrm{b}} = \left(\Omega^{\#}\right)^{-1}$; clearly, $\Omega^{\mathrm{b}} X_H\left(\mathbf{z}\right) = \nabla H\left(\mathbf{z}\right)$, and the dynamical equations transform into

$$\Omega^{\mathrm{b}}\dot{\mathbf{z}} = \nabla H\left(\mathbf{z}\right). \tag{43}$$

**2.8 Complex Manifold and Dynamical Equations**

An n-dimensional complex manifold is a complex space $\chi$ with the following properties: a) $\chi$ is a Hausdorff space; b) $\chi$ has a countable basis; and c) $\chi$ is equipped with an n-dimensional complex structure (Fritzsche & Grauert. 2002). If the dynamical equations of a particle in the potential field of a rotating asteroid can be expressed on the complex manifold, then the equations can be studied on the complex manifold with the theories of complex manifolds.

**2.8.1 Dynamical Equation on the Kähler Manifold**

A complex manifold with a symplectic structure is a Kähler manifold (Berndt. 1998). If we write

$$\boldsymbol{\Gamma} = \mathbf{q} + i\mathbf{p} = \mathbf{r} + im\left(\dot{\mathbf{r}} + \boldsymbol{\omega}\times\mathbf{r}\right), \tag{44}$$

then it is easy to show that the dynamical equations of a particle near a rotating asteroid expressed on the Kähler manifold can be written as (Marsden & Ratiu. 1999)

$$\boldsymbol{\Gamma} = -2i\frac{\partial H}{\partial \overline{\boldsymbol{\Gamma}}}, \tag{45}$$

where $\frac{\partial}{\partial \overline{\boldsymbol{\Gamma}}} \triangleq \frac{1}{2}\left(\frac{\partial}{\partial \mathbf{q}} + i\frac{\partial}{\partial \mathbf{p}}\right)$.

### 2.8.2 Dynamical Equations on a Complex Manifold

Let

$$\begin{cases} z_1 = x + iv_x \\ z_2 = y + iv_y \\ z_3 = z + iv_z \end{cases}; \tag{46}$$

then the dynamical equations can be transformed into the form of several complex variables, and the motion of a particle in the potential field of a rotating asteroid can be studied using multi-complex function theory.

It is easy to show that

$$\begin{cases} x = \dfrac{z_1 + \bar{z}_1}{2} \\ v_x = \dfrac{z_1 - \bar{z}_1}{2i} \end{cases} \quad \begin{cases} y = \dfrac{z_2 + \bar{z}_2}{2} \\ v_y = \dfrac{z_2 - \bar{z}_2}{2i} \end{cases} \quad \begin{cases} z = \dfrac{z_3 + \bar{z}_3}{2} \\ v_z = \dfrac{z_3 - \bar{z}_3}{2i} \end{cases}. \tag{47}$$

The dynamical equations of the particle in multi-complex analysis form are

$$\frac{1}{2i}\frac{d}{dt}\begin{bmatrix} z_1 - \bar{z}_1 \\ z_2 - \bar{z}_2 \\ z_3 - \bar{z}_3 \end{bmatrix} + \frac{1}{i}\hat{\boldsymbol{\omega}}\begin{bmatrix} z_1 - \bar{z}_1 \\ z_2 - \bar{z}_2 \\ z_3 - \bar{z}_3 \end{bmatrix} + \frac{1}{2}\hat{\boldsymbol{\omega}}\hat{\boldsymbol{\omega}}\begin{bmatrix} z_1 + \bar{z}_1 \\ z_2 + \bar{z}_2 \\ z_3 + \bar{z}_3 \end{bmatrix} + \frac{1}{2}\hat{\boldsymbol{\tau}}\begin{bmatrix} z_1 + \bar{z}_1 \\ z_2 + \bar{z}_2 \\ z_3 + \bar{z}_3 \end{bmatrix} + \frac{1}{2}\begin{bmatrix} \dfrac{\partial U(\mathbf{z},\bar{\mathbf{z}})}{\partial z_1} + \dfrac{\partial U(\mathbf{z},\bar{\mathbf{z}})}{\partial \bar{z}_1} \\ \dfrac{\partial U(\mathbf{z},\bar{\mathbf{z}})}{\partial z_2} + \dfrac{\partial U(\mathbf{z},\bar{\mathbf{z}})}{\partial \bar{z}_2} \\ \dfrac{\partial U(\mathbf{z},\bar{\mathbf{z}})}{\partial z_3} + \dfrac{\partial U(\mathbf{z},\bar{\mathbf{z}})}{\partial \bar{z}_3} \end{bmatrix} = 0 \tag{48}$$

or

$$\frac{d}{dt}\begin{bmatrix} z_1 - \bar{z}_1 \\ z_2 - \bar{z}_2 \\ z_3 - \bar{z}_3 \end{bmatrix} + 2\hat{\boldsymbol{\omega}}\begin{bmatrix} z_1 - \bar{z}_1 \\ z_2 - \bar{z}_2 \\ z_3 - \bar{z}_3 \end{bmatrix} + i\hat{\boldsymbol{\omega}}\hat{\boldsymbol{\omega}}\begin{bmatrix} z_1 + \bar{z}_1 \\ z_2 + \bar{z}_2 \\ z_3 + \bar{z}_3 \end{bmatrix} + i\hat{\boldsymbol{\tau}}\begin{bmatrix} z_1 + \bar{z}_1 \\ z_2 + \bar{z}_2 \\ z_3 + \bar{z}_3 \end{bmatrix} + i\begin{bmatrix} \dfrac{\partial U(\mathbf{z},\bar{\mathbf{z}})}{\partial z_1} + \dfrac{\partial U(\mathbf{z},\bar{\mathbf{z}})}{\partial \bar{z}_1} \\ \dfrac{\partial U(\mathbf{z},\bar{\mathbf{z}})}{\partial z_2} + \dfrac{\partial U(\mathbf{z},\bar{\mathbf{z}})}{\partial \bar{z}_2} \\ \dfrac{\partial U(\mathbf{z},\bar{\mathbf{z}})}{\partial z_3} + \dfrac{\partial U(\mathbf{z},\bar{\mathbf{z}})}{\partial \bar{z}_3} \end{bmatrix} = 0 \tag{49}$$

where $\mathbf{z} = \begin{bmatrix} z_1 \\ z_2 \\ z_3 \end{bmatrix}$ and $\bar{\mathbf{z}} = \begin{bmatrix} \bar{z}_1 \\ \bar{z}_2 \\ \bar{z}_3 \end{bmatrix}$.

Using the effective potential, the dynamical equations of the particle in multi-complex analysis form can be expressed as

$$\frac{1}{2i}\frac{d}{dt}\begin{bmatrix} z_1-\bar{z}_1 \\ z_2-\bar{z}_2 \\ z_3-\bar{z}_3 \end{bmatrix}+\frac{1}{i}\hat{\boldsymbol{\omega}}\begin{bmatrix} z_1-\bar{z}_1 \\ z_2-\bar{z}_2 \\ z_3-\bar{z}_3 \end{bmatrix}+\frac{1}{2}\hat{\boldsymbol{\tau}}\begin{bmatrix} z_1+\bar{z}_1 \\ z_2+\bar{z}_2 \\ z_3+\bar{z}_3 \end{bmatrix}+\frac{1}{2}\begin{bmatrix} \frac{\partial V(\mathbf{z},\bar{\mathbf{z}})}{\partial z_1}+\frac{\partial V(\mathbf{z},\bar{\mathbf{z}})}{\partial \bar{z}_1} \\ \frac{\partial V(\mathbf{z},\bar{\mathbf{z}})}{\partial z_2}+\frac{\partial V(\mathbf{z},\bar{\mathbf{z}})}{\partial \bar{z}_2} \\ \frac{\partial V(\mathbf{z},\bar{\mathbf{z}})}{\partial z_3}+\frac{\partial V(\mathbf{z},\bar{\mathbf{z}})}{\partial \bar{z}_3} \end{bmatrix}=0 \qquad (50)$$

or

$$\frac{d}{dt}\begin{bmatrix} z_1-\bar{z}_1 \\ z_2-\bar{z}_2 \\ z_3-\bar{z}_3 \end{bmatrix}+2\hat{\boldsymbol{\omega}}\begin{bmatrix} z_1-\bar{z}_1 \\ z_2-\bar{z}_2 \\ z_3-\bar{z}_3 \end{bmatrix}+i\hat{\boldsymbol{\tau}}\begin{bmatrix} z_1+\bar{z}_1 \\ z_2+\bar{z}_2 \\ z_3+\bar{z}_3 \end{bmatrix}+i\begin{bmatrix} \frac{\partial V(\mathbf{z},\bar{\mathbf{z}})}{\partial z_1}+\frac{\partial V(\mathbf{z},\bar{\mathbf{z}})}{\partial \bar{z}_1} \\ \frac{\partial V(\mathbf{z},\bar{\mathbf{z}})}{\partial z_2}+\frac{\partial V(\mathbf{z},\bar{\mathbf{z}})}{\partial \bar{z}_2} \\ \frac{\partial V(\mathbf{z},\bar{\mathbf{z}})}{\partial z_3}+\frac{\partial V(\mathbf{z},\bar{\mathbf{z}})}{\partial \bar{z}_3} \end{bmatrix}=0\,. \qquad (51)$$

The effective potential is written as

$$V(\mathbf{z},\bar{\mathbf{z}})=U(\mathbf{z},\bar{\mathbf{z}})-\frac{1}{8}\left|\hat{\boldsymbol{\omega}}(\mathbf{z}+\bar{\mathbf{z}})\right|^2=U(\mathbf{z},\bar{\mathbf{z}})-\frac{1}{8}\left(\hat{\boldsymbol{\omega}}(\mathbf{z}+\bar{\mathbf{z}})\right)^2. \qquad (52)$$

The integral of the relative energy becomes

$$H=U(\mathbf{z},\bar{\mathbf{z}})-\frac{1}{8}\left|\mathbf{z}-\bar{\mathbf{z}}\right|^2-\frac{1}{8}\left|\hat{\boldsymbol{\omega}}(\mathbf{z}+\bar{\mathbf{z}})\right|^2=U(\mathbf{z},\bar{\mathbf{z}})-\frac{1}{8}(\mathbf{z}-\bar{\mathbf{z}})^2-\frac{1}{8}\left(\hat{\boldsymbol{\omega}}(\mathbf{z}+\bar{\mathbf{z}})\right)^2; \qquad (53)$$

using the effective potential, the integral of the relative energy is given by

$$H=V(\mathbf{z},\bar{\mathbf{z}})-\frac{1}{8}\left|\mathbf{z}-\bar{\mathbf{z}}\right|^2=V(\mathbf{z},\bar{\mathbf{z}})-\frac{1}{8}(\mathbf{z}-\bar{\mathbf{z}})^2. \qquad (54)$$

If the attitude motion of the asteroid is uniformly rotating, then $\hat{\boldsymbol{\tau}}=0_{3\times3}$ is a $3\times3$ zero matrix. The dynamical equations take the form

$$\frac{1}{2i}\frac{d}{dt}\begin{bmatrix} z_1-\bar{z}_1 \\ z_2-\bar{z}_2 \\ z_3-\bar{z}_3 \end{bmatrix}+\frac{1}{i}\hat{\boldsymbol{\omega}}\begin{bmatrix} z_1-\bar{z}_1 \\ z_2-\bar{z}_2 \\ z_3-\bar{z}_3 \end{bmatrix}+\frac{1}{2}\hat{\boldsymbol{\omega}}\hat{\boldsymbol{\omega}}\begin{bmatrix} z_1+\bar{z}_1 \\ z_2+\bar{z}_2 \\ z_3+\bar{z}_3 \end{bmatrix}+\frac{1}{2}\begin{bmatrix} \frac{\partial U(\mathbf{z},\bar{\mathbf{z}})}{\partial z_1}+\frac{\partial U(\mathbf{z},\bar{\mathbf{z}})}{\partial \bar{z}_1} \\ \frac{\partial U(\mathbf{z},\bar{\mathbf{z}})}{\partial z_2}+\frac{\partial U(\mathbf{z},\bar{\mathbf{z}})}{\partial \bar{z}_2} \\ \frac{\partial U(\mathbf{z},\bar{\mathbf{z}})}{\partial z_3}+\frac{\partial U(\mathbf{z},\bar{\mathbf{z}})}{\partial \bar{z}_3} \end{bmatrix}=0 \qquad (55)$$

or

$$\frac{d}{dt}\begin{bmatrix} z_1-\bar{z}_1 \\ z_2-\bar{z}_2 \\ z_3-\bar{z}_3 \end{bmatrix}+2\hat{\boldsymbol{\omega}}\begin{bmatrix} z_1-\bar{z}_1 \\ z_2-\bar{z}_2 \\ z_3-\bar{z}_3 \end{bmatrix}+i\hat{\boldsymbol{\omega}}\hat{\boldsymbol{\omega}}\begin{bmatrix} z_1+\bar{z}_1 \\ z_2+\bar{z}_2 \\ z_3+\bar{z}_3 \end{bmatrix}+i\begin{bmatrix} \frac{\partial U(\mathbf{z},\bar{\mathbf{z}})}{\partial z_1}+\frac{\partial U(\mathbf{z},\bar{\mathbf{z}})}{\partial \bar{z}_1} \\ \frac{\partial U(\mathbf{z},\bar{\mathbf{z}})}{\partial z_2}+\frac{\partial U(\mathbf{z},\bar{\mathbf{z}})}{\partial \bar{z}_2} \\ \frac{\partial U(\mathbf{z},\bar{\mathbf{z}})}{\partial z_3}+\frac{\partial U(\mathbf{z},\bar{\mathbf{z}})}{\partial \bar{z}_3} \end{bmatrix}=0. \qquad (56)$$

Using the effective potential, the dynamical equations of a particle near the uniformly rotating asteroid can be expressed in multi-complex analysis form as

$$\frac{1}{2i}\frac{d}{dt}\begin{bmatrix} z_1-\bar{z}_1 \\ z_2-\bar{z}_2 \\ z_3-\bar{z}_3 \end{bmatrix}+\frac{1}{i}\hat{\boldsymbol{\omega}}\begin{bmatrix} z_1-\bar{z}_1 \\ z_2-\bar{z}_2 \\ z_3-\bar{z}_3 \end{bmatrix}+\frac{1}{2}\begin{bmatrix} \frac{\partial V(\mathbf{z},\bar{\mathbf{z}})}{\partial z_1}+\frac{\partial V(\mathbf{z},\bar{\mathbf{z}})}{\partial \bar{z}_1} \\ \frac{\partial V(\mathbf{z},\bar{\mathbf{z}})}{\partial z_2}+\frac{\partial V(\mathbf{z},\bar{\mathbf{z}})}{\partial \bar{z}_2} \\ \frac{\partial V(\mathbf{z},\bar{\mathbf{z}})}{\partial z_3}+\frac{\partial V(\mathbf{z},\bar{\mathbf{z}})}{\partial \bar{z}_3} \end{bmatrix}=0 \qquad (57)$$

or

$$\frac{d}{dt}\begin{bmatrix} z_1-\bar{z}_1 \\ z_2-\bar{z}_2 \\ z_3-\bar{z}_3 \end{bmatrix}+2\hat{\boldsymbol{\omega}}\begin{bmatrix} z_1-\bar{z}_1 \\ z_2-\bar{z}_2 \\ z_3-\bar{z}_3 \end{bmatrix}+i\begin{bmatrix} \frac{\partial V(\mathbf{z},\bar{\mathbf{z}})}{\partial z_1}+\frac{\partial V(\mathbf{z},\bar{\mathbf{z}})}{\partial \bar{z}_1} \\ \frac{\partial V(\mathbf{z},\bar{\mathbf{z}})}{\partial z_2}+\frac{\partial V(\mathbf{z},\bar{\mathbf{z}})}{\partial \bar{z}_2} \\ \frac{\partial V(\mathbf{z},\bar{\mathbf{z}})}{\partial z_3}+\frac{\partial V(\mathbf{z},\bar{\mathbf{z}})}{\partial \bar{z}_3} \end{bmatrix}=0. \qquad (58)$$

## 2.9 Cohomology and Dynamical Equations

Using cohomology theory, the dynamical equations of a particle near the uniformly rotating asteroid can be expressed simply and beautifully. The Hodge star operator is a significant linear map which defined on the exterior algebra of a finite-dimensional oriented inner product space. Using the Hodge star operator, the dynamical equation of a particle near the uniformly rotating asteroid is equivalent to the duality of two 1-vectors.

Let us write the kinetic energy as

$$T = \frac{1}{2}\left(\dot{\mathbf{q}} + \boldsymbol{\omega} \times \mathbf{q}\right) \cdot \left(\dot{\mathbf{q}} + \boldsymbol{\omega} \times \mathbf{q}\right). \tag{59}$$

Let

$$\begin{cases} \mathbf{T}^1 = \dfrac{\partial T}{\partial \dot{\mathbf{q}}} dt \\ \mathbf{U}^1 = -U d\mathbf{q} \end{cases}. \tag{60}$$

Apply the Hodge star operator $*$ and the differential operator $d$ to Eq. (60).

$$d * \mathbf{T}^1 = \frac{d}{dt}\left(\frac{\partial T}{\partial \dot{\mathbf{q}}}\right) dt \wedge d\dot{\mathbf{q}} \wedge d\mathbf{q} \tag{61}$$

$$d * \mathbf{U}^1 = -\frac{\partial U}{\partial \mathbf{q}} d\mathbf{q} \wedge dt \wedge d\dot{\mathbf{q}} \tag{62}$$

The dynamical equations of the particle can then be written as

$$d * \mathbf{T}^1 = d * \mathbf{U}^1. \tag{63}$$

If the kinetic energy and the effective potential can be expressed as

$$\begin{cases} T_V = \dfrac{1}{2}\dot{\mathbf{q}} \cdot \dot{\mathbf{q}} \\ V\left(\mathbf{q}\right) = -\dfrac{1}{2}\left(\boldsymbol{\omega} \times \mathbf{q}\right) \cdot \left(\boldsymbol{\omega} \times \mathbf{q}\right) + U\left(\mathbf{q}\right) \end{cases} \tag{64}$$

and we define

$$\begin{cases} \mathbf{T}_V^1 = \dfrac{\partial T_V}{\partial \dot{\mathbf{q}}} dt \\ \mathbf{V}^1 = -V d\mathbf{q} \end{cases}, \tag{65}$$

then, using the Hodge star operator $*$ and the differential operator $d$

$$d*\mathbf{T}_V^1=\frac{d}{dt}\left(\frac{\partial T_V}{\partial\dot{\mathbf{q}}}\right)dt\wedge d\dot{\mathbf{q}}\wedge d\mathbf{q}, \tag{66}$$

$$d*\mathbf{V}^1=-\frac{\partial V}{\partial\mathbf{q}}d\mathbf{q}\wedge dt\wedge d\dot{\mathbf{q}}. \tag{67}$$

The dynamical equations of the particle can then be written as

$$d*\mathbf{T}_V^1=d*\mathbf{V}^1. \tag{68}$$

Eq. (63) and Eq. (68) are the dynamical equations of a particle in the potential field of a rotating asteroid expressed in cohomology form.

## 3 Summary of the Dynamical Equations

The classical form of the dynamical equations includes the second-order derivative of the position vector of the particle, which is the radius vector from the asteroid's centre of mass to the particle. There are 2 classical forms of the dynamical equations; one is written in terms of the potential, while the other uses the effective potential. For the uniformly rotating asteroid, the dynamical equations can be simplified. There are 4 scalar forms of the general dynamical equations for a particle near a rotating asteroid presented here, using the potential or the effective potential and using the arbitrary body-fixed frame or the special body-fixed frame. If the dynamical equations are expressed in the special body-fixed frame, then the effective potential is simplified to Eq. (15), the Jacobi integral is simplified to Eq. (16), and the Lagrange function is reduced to Eq. (17). In addition, the asymptotic surface of the effective potential expressed in the special body-fixed frame is a circular cylindrical surface.

The dynamical equations of the particle in coefficient-matrix form appear as a first-order ordinary differential equation, and dynamical systems theory can be easily applied to this form of the dynamical equations to study the orbital dynamical system, the bifurcation, and the chaotic motion of the orbital motion of a particle near a rotating asteroid.

The dynamical equations of a particle in the potential field of a rotating asteroid expressed in the Lagrange form regard the motion of the particle as the autonomous curve of the functional $\Phi = \int_{t_0}^{t_1} L dt$. To express the dynamical equations of a particle near a rotating asteroid in phase space, one can utilise the dynamical equations in Hamilton form. The dynamical equations can also be given in symplectic form with the Hamiltonian vector field on the symplectic manifold. If the asteroid rotates uniformly, then the Jacobi integral is time invariant, and the symplectic geometric algorithm can be used to integrate the dynamical equations. Using the Poisson bracket, the dynamical equations of a particle in the potential field of a rotating asteroid can be rewritten in Poisson-bracket form on the symplectic manifold. In addition, the dynamical equations of a particle near a rotating asteroid can be expressed on the Poisson manifold with the entire smooth Hamiltonian vector field. This leads to a novel method of studying the motion of a particle near a rotating asteroid on the Poisson manifold.

The dynamical equations on the Kähler manifold take on a simplified form and appear as a first-order complex differential equation. The dynamical equations on the other complex manifold considered here exhibit a complicated form and lead to a novel method of studying the motion of a particle near a rotating asteroid using multi-complex function theory.

The dynamical equations can be expressed simply and beautifully in cohomology form; the Hodge star operator and the differential operator are applied to express the motion of a particle near a rotating asteroid, and this form of the dynamical equations allows the conclusions of the cohomology group to be applied to the study of the motion of a particle in the potential field of a rotating asteroid.

Nine types of dynamical models of a particle near a rotating asteroid are presented: the scalar form, the coefficient-matrix form, the Lagrange form, the Hamilton form, the symplectic form, the Poisson-bracket form, the Poisson form, the complex form and the cohomology form. The scalar forms of dynamical models in Sec. 2.2.1 and 2.2.2 are suitable to all of the irregular asteroids, while the scalar forms of dynamical models in Sec. 2.2.3 are only suitable to uniformly rotating irregular asteroids. Besides, the coefficient-matrix form, the Lagrange form, the Hamilton form, the symplectic form, the Poisson-bracket form and the Poisson form of dynamical models are suitable to all of the irregular asteroids. The cohomology form of dynamical model is only suitable to uniformly rotating irregular asteroids. About the complex forms of dynamical models, Eq. (45), (48), (49), (50) and (51) are suitable to all of the irregular asteroids, while Eq. (55), (56), (57) and (58) are only suitable to uniformly rotating irregular asteroids.

The difference among different formalisms is comprised of: the characteristic of the equation; which manifold that the solution of the equation is in; the scope of application that for uniformly rotating or variable-velocity rotation asteroids; the formalism of the potential functions; available or unavailable for numerical calculation etc. The characteristic of the equation includes classical, coefficient-matrix, Lagrange, Hamilton, symplectic, Poisson-bracket, Poisson, complex and cohomology form; the formalism of the potential function embodies potential or effective potential. Table 1 shows a comparison among the various forms of the dynamical equations of a particle in the potential field of a rotating asteroid. Each form of the dynamical equations can be written with 2 types of potential-field functions: the potential and the effective potential. The dynamical equations of a particle in the potential field of the uniformly rotating asteroid can be obtained from the dynamical equations in the potential field of a variable-velocity rotating asteroid and have a simpler form. The collective advantage of classical, coefficient-matrix, Lagrange, Hamilton, symplectic and complex forms is available for numerical calculation. The collective advantage of Hamilton, symplectic and Kähler forms is available for symplectic numerical calculation. The disadvantage of Poisson-bracket, Poisson, and cohomology forms is unavailable for directly numerical calculation; besides, the Poisson-bracket and Poisson form of equations can be used for numerical calculation when it is transformed into Hamilton or symplectic forms while the cohomology form of equation can be used for numerical calculation when it is transformed into Lagrange form.

Solutions of different equations are in different manifolds and have different immanent behaviors. The solution of the scalar form or the coefficient-matrix form is in the body-fixed frame, which can be used to calculate the orbits of the particle orbiting a rotating asteroid with numerical method easily. The solution of the Hamilton form or the symplectic form is in the symplectic manifold and can be used to calculate the orbits of the particle orbiting the uniformly rotating asteroid with symplectic numerical method; besides, the conservation of the Jacobi integral is has good numerical phenomenon. The solution of the complex form is in the complex manifold; specially, the solution of the complex form which is on the Kähler Manifold is the corresponding solution with symplectic behaviors in the complex manifold. The solution in the cohomology form is the trajectory which satisfies the Hodge star operator equation.

Table 1 Comparison of Different Forms of the Dynamical Equations of a Particle near an Asteroid

| Equation Number | Equation Characteristic | Manifold | Uniformly Rotating or Variable-Velocity Rotation | Potential or Effective potential |
|---|---|---|---|---|
| Eq. (1)/ Eq. (12)/ Eq. (14) | Classical Form | 3-dimensional Euclidean Space | Variable Velocity | Potential |
| Eq. (6)/ Eq. (13)/ Eq. (18) | Classical Form | 3-dimensional Euclidean Space | Variable Velocity | Effective potential |
| Eq. (9)/ Eq. (19) | Classical Form | 3-dimensional Euclidean Space | Uniformly Rotating | Potential |
| Eq. (8)/ Eq. (20) | Classical Form | 3-dimensional Euclidean Space | Uniformly Rotating | Effective potential |
| Eq. (22) | Coefficient-Matrix Form | 6-dimensional Space | Variable Velocity | Effective potential |
| Eq. (26) | Lagrange Form | Phase Space | Variable Velocity | Potential / Effective potential |
| Eq. (31) | Hamilton Form | Phase Space | Variable Velocity | Potential / Effective potential |
| Eq. (33)/ Eq. (34) | Symplectic Form | Symplectic Manifold | Variable Velocity | Potential / Effective potential |
| Eq. (36) | Poisson-Bracket Form | Symplectic Manifold | Variable Velocity | Potential / Effective potential |
| Eq. (41)/ Eq. (42)/ Eq. (43) | Poisson Form | Poisson Manifold | Variable Velocity | Potential / Effective potential |
| Eq. (45) | Complex Form | Kähler Manifold & Complex Manifold | Variable Velocity | Potential / Effective potential |
| Eq. (48)/ Eq. (49) | Complex Form | Complex Manifold | Variable Velocity | Potential |
| Eq. (50)/ Eq. (51) | Complex Form | Complex Manifold | Variable Velocity | Effective potential |
| Eq. (55)/ Eq. (56) | Complex Form | Complex Manifold | Uniformly Rotating | Potential |
| Eq. (57)/ Eq. (58) | Complex Form | Complex Manifold | Uniformly Rotating | Effective potential |
| Eq. (63)/ Eq. (68) | Cohomology Form | Phase Space | Uniformly Rotating | Potential / Effective potential |

## 4 Orbital Dynamics Close to Asteroids 216 Kleopatra and 1620 Geographos

In this section, the orbital dynamics equations around a rotating asteroid are applied to 216 Kleopatra and 1620 Geographos. We used the coefficient-matrix form of dynamical models to simulate the orbital dynamics close to asteroids 216 Kleopatra and 1620 Geographos, because this dynamics model is easy and suitable for programming. The methodology for the application of the coefficient-matrix form of dynamical models to simulate the orbital dynamics close to asteroids is: Let the unit vector $\mathbf{e_z}$ is defined by $\boldsymbol{\omega} = \omega \mathbf{e_z}$. Integrate the first-order ordinary differential equation Eq. (22), which is $\dot{\mathbf{X}} = \mathbf{AX} + \mathbf{B}(\mathbf{X})$, where $\mathbf{X} = \begin{bmatrix} \mathbf{r} \\ \mathbf{v} \end{bmatrix}$, $\mathbf{A} = \begin{pmatrix} \mathbf{I}_{3\times3} & \mathbf{0}_{3\times3} \\ -2\hat{\boldsymbol{\omega}} & -\hat{\boldsymbol{\tau}} \end{pmatrix}$, $\mathbf{B} = \begin{pmatrix} \mathbf{0}_{3\times3} \\ -\nabla V(\mathbf{r}) \end{pmatrix}$ and $\nabla V(\mathbf{r}) = \nabla U(\mathbf{r}) - \omega^2 (x + y)$. $\nabla U(\mathbf{r})$ is calculated by the polyhedron model (Werner 1994; Werner and Scheeres 1996) using the data from radar observations (Neese 2004).

The rotation period of 216 Kleopatra is 5.385 h and with overall dimensions of $217 \times 94 \times 81$ km (Ostro et al. 2000); in addition, the estimated bulk density is 3.6 $\mathrm{g \cdot cm^{-3}}$ (Descamps et al. 2010). The physical model of 216 Kleopatra that we used here was calculated with radar observations using the polyhedral model with 2048 vertices and 4096 faces (Neese 2004). The unit vector $\mathbf{e_z}$ of the body-fixed frame is defined by $\boldsymbol{\omega} = \omega \mathbf{e_z}$, the initial position of the particle in the body-fixed frame is $\mathbf{r} = [\ -92863.5, 49248.6, 21413.9]\mathrm{m}$ and the initial velocity is $\mathbf{v} = [42.247, 71.958, -3.468]\mathrm{m \cdot s^{-1}}$. The total flight time of the particle is 27h42min. Figure 1 shows the orbit of the particle around the asteroid 216 Kleopatra with the initial position and initial velocity in the body-fixed frame.

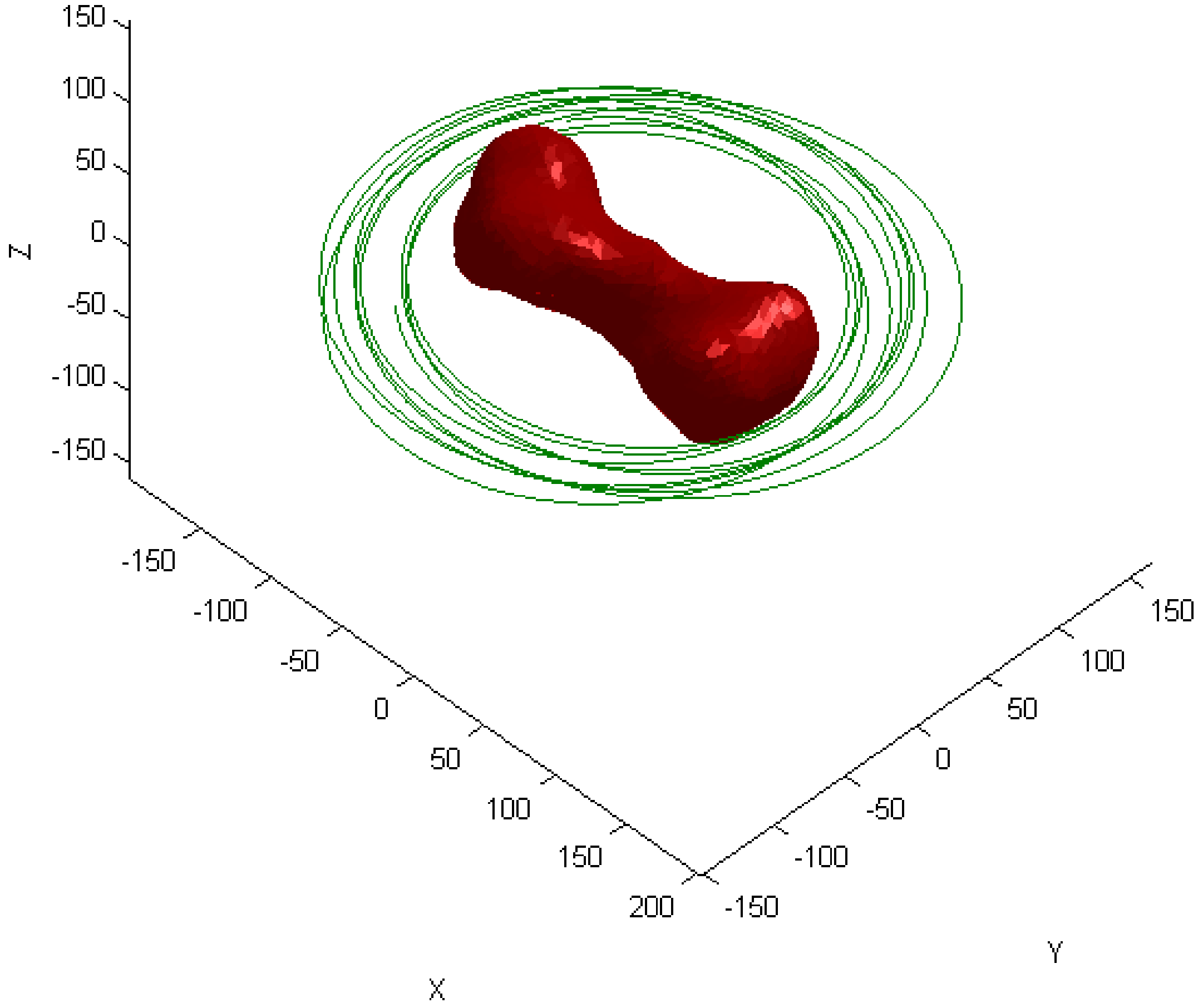

Figure 1. An orbit of the particle around the asteroid 216 Kleopatra (km)

The rotation period of 1620 Geographos is 5.222h (Ryabova 2002), the estimated bulk density is $2.0 \ \ \mathrm{g \cdot cm^{-3}}$ (Hudson & Ostro 1999), and with overall dimensions of $(5.0 \times 2.0 \times 2.1) \pm 0.15$ km (Hudson & Ostro 1999). The physical model of 1620 Geographos that we used here was calculated with radar observations using the polyhedral model with 8192 vertices and 16380 faces (Neese 2004). The initial position of the particle in the body-fixed frame is $\mathbf{r} = [\ 139995.2, -600.4, -7053.3]\mathrm{m}$ and the initial velocity is $\mathbf{v} = [-0.8944, -4.3202, 0.42324]\,\mathrm{m \cdot s^{-1}}$. The total flight time of the particle is 10h6min. Figure 2a shows the orbit of the particle around the asteroid 1620 Geographos with the initial position and initial velocity in the body-fixed frame, while figure 2b shows the orbit in the inertia frame. From figure 2, one can know the particle with this orbital initial parameter will leave asteroid 1620 Geographos.

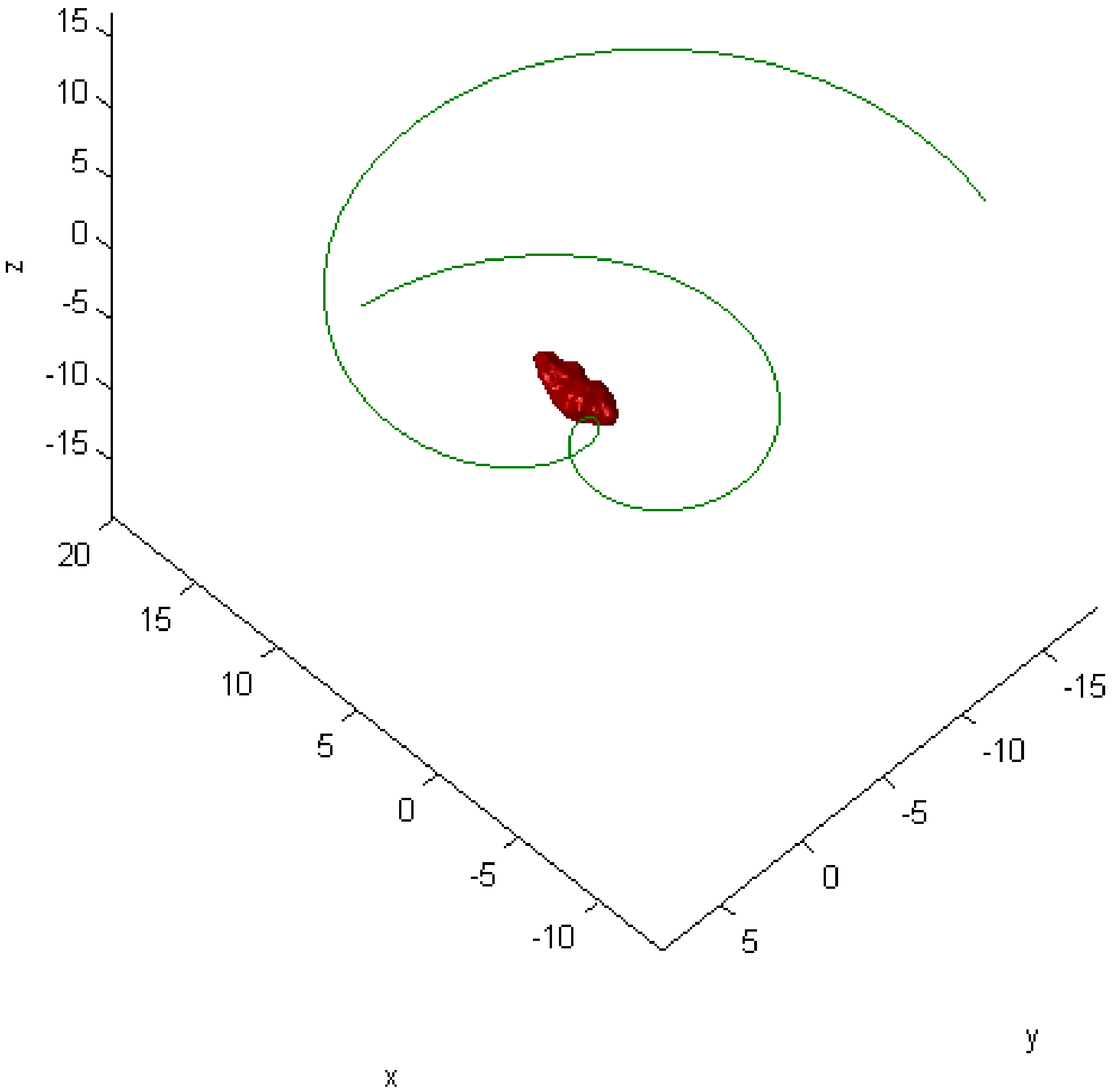

Figure 2a. An orbit of the particle around the asteroid 1620 Geographos in the body-fixed frame (km)

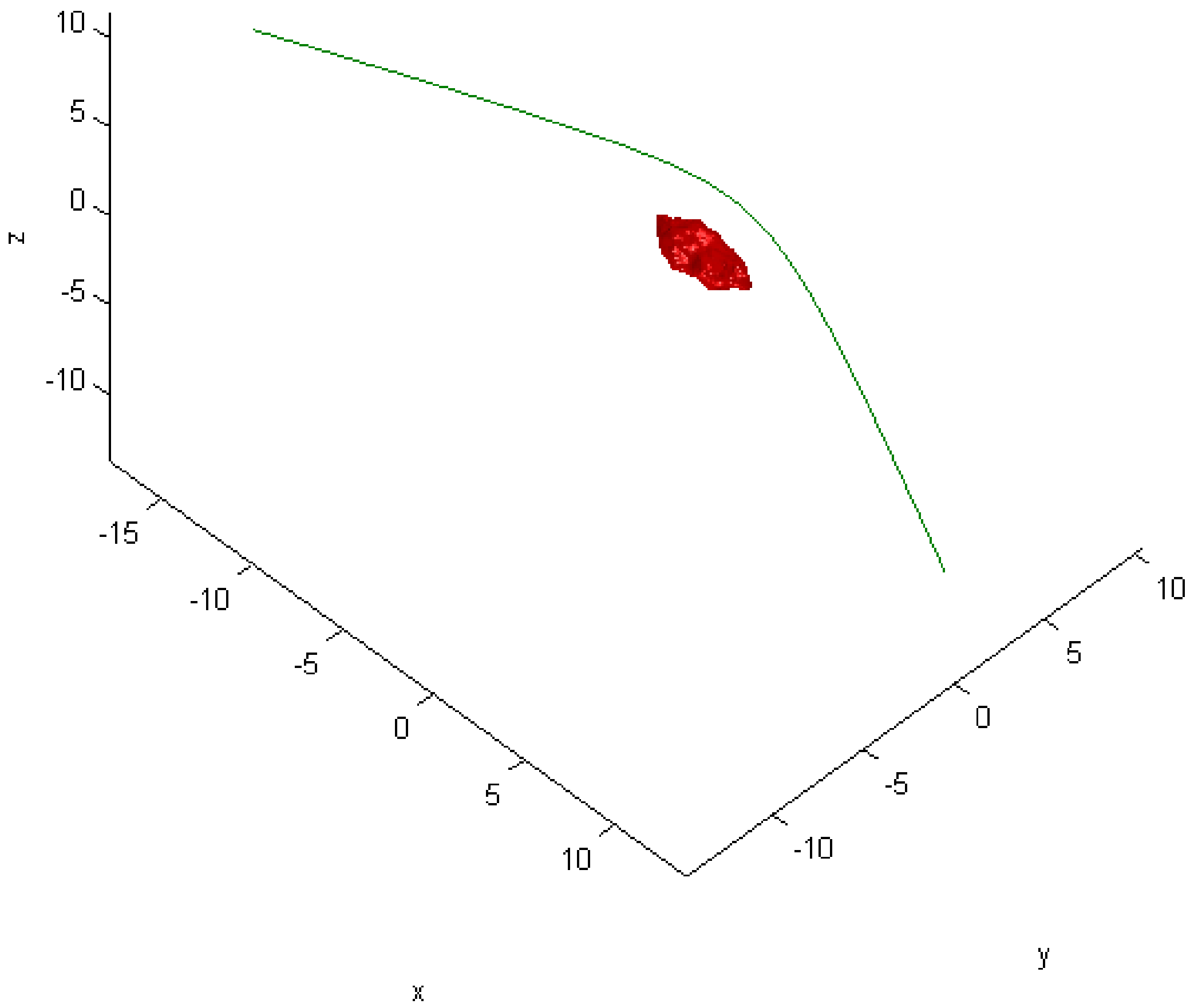

Figure 2b. An orbit of the particle around the asteroid 1620 Geographos in the inertia frame (km)

## 5 Conclusions

In addition to the classical form of the dynamical equations, 9 new forms of the dynamical equations of a particle orbiting a rotating asteroid have been derived and presented: the scalar form, the coefficient-matrix form, the Lagrange form, the Hamilton form, the symplectic form, the Poisson-bracket form, the Poisson form, the cohomology form, and the dynamical equations on the Kähler manifold and another complex manifold. Novel forms of the effective potential and the Jacobi integral have also been presented.

The dynamical equations in scalar form with the potential and the effective potential in the arbitrary body-fixed frame and the special body-fixed frame were presented and studied, and the simplified forms of the effective potential and the Jacobi integral were given. The dynamical equations in coefficient-matrix form have been derived and shown to take the form of a first-order ordinary differential equation; expressing the dynamical equations in this form can aid in the study of the dynamical system, the bifurcation, and the chaotic motion of the orbital dynamics of a particle near a rotating asteroid.

The dynamical equations in symplectic form and Poisson-bracket form represent the motion of the particle expressed on the symplectic manifold, while the dynamical equations in Poisson form represent the motion of the particle expressed on the Poisson manifold. The motion of a particle in the potential field of a rotating asteroid can also be expressed on a complex manifold, including the Kähler manifold. The dynamical equations on the Kähler manifold take on a simplified form and appear as a first-order complex differential equation. The dynamical equations on the other complex manifold considered here have a complicated form and lead to a novel method of studying the motion of a particle near a rotating asteroid using multi-complex function theory. Using the Hodge star operator and the differential operator, the dynamical equations of the particle can be expressed in cohomology form; this form of the dynamical equations looks simple and beautiful, and it allows the conclusions of the cohomology group to be applied to the study of the motion of a particle in the potential field of a rotating asteroid.

## Acknowledgements

This research was supported by the National Natural Science Foundation of China (No. 11072122), the State Key Laboratory Foundation of Astronautic Dynamics (No. 2012ADL0202), and the National Basic Research Program of China (973 Program, 2012CB720000).